# Zero-dimensional Analysis of Thermal Choking in Arbitrary Duct with Application to Dual-mode Ramjet Combustor

N. Ananthkrishnan[1] and K. Sreesankar[2]

*Yanxiki Tech, 152 Clover Parkview, Koregaon Park, Pune 411001, India*

**Thermal choking in combustor ducts is a matter of great interest to researchers involved in the design and analysis of combustors. In case of scramjet combustors, a thermally choked state is undesirable whereas for a dual-mode combustor in the ram mode it is essential. Since real-life combustors feature volume heat addition due to combustion, area variation, wall heat loss and frictional losses, all of these effects must be included in the analysis. The standard Rayleigh flow assumes a constant-area duct with no frictional effects and a single heat addition/loss term without distinguishing between combustion and wall heat transfer effects. Instead, a generalized Rayleigh flow analysis is employed here which can account for all the above effects and also provide for different fuel injection rates and mixing efficiency. The "forward" and "inverse" heat transfer problems are stated and solutions are compared between Rayleigh flow in constant-area and variable-area ducts. The analysis is checked against a scramjet combustor in the literature and then used to study the occurrence of thermal choking in a dual-mode combustor geometry. The possibility of thermal choking and the location of the choking event can be easily detected by this method. The shift in the thermal choking location with variation in the duct heat and mass transfer state can be easily evaluated. In this manner, the present method provides a quick and accurate tool for preliminary design and analysis of ramjet and scramjet combustors.**

## Nomenclature

$A$ = area

$AR$ = area ratio, $A_2/A_1$

$\bar{A}_w$ = non-dimensional wetted area, $A_w/A_1$

---

[1] Technical Consultant, Associate Fellow AIAA; akn.korea.19@gmail.com, akn@aero.iitb.ac.in

[2] Technical Consultant; sreeshankark@gmail.com

| | | |
|---|---|---|
| $C_f$ | = | friction coefficient |
| $F$ | = | vector of nonlinear functions |
| $f$ | = | fuel-air mass ratio |
| $M$ | = | Mach number |
| $p$ | = | pressure |
| $\dot{Q}$ | = | heat addition rate |
| $\bar{\dot{Q}}_c$ | = | non-dimensional heat addition rate due to combustion, $\dot{Q}_c/\rho_1 U_1^3 A_1$ |
| $\bar{\dot{Q}}_w$ | = | non-dimensional heat addition rate due to wall heat transfer, $\dot{Q}_w/\rho_1 U_1^3 A_1$ |
| $R$ | = | gas constant for air |
| $\bar{R}$ | = | $R_2/R_1$ |
| $r$ | = | density ratio, $\rho_2/\rho_1$ |
| $T$ | = | temperature |
| $U$ | = | velocity |
| $u$ | = | $U_2/U_1$ |
| $\gamma$ | = | ratio of specific heats |
| $\bar{\gamma}$ | = | $\gamma_2/\gamma_1$ |
| $\rho$ | = | density |

*Subscripts*

| | | |
|---|---|---|
| $c$ | = | combustion |
| $w$ | = | wall or wetted |
| $0$ | = | total or stagnation value |
| $1,2$ | = | inlet and outlet station, respectively |

## I. Introduction

The analysis of frictionless flow in a constant-area duct with heat addition or removal, called *Rayleigh flow*, is a widely discussed topic in several textbooks [1, pp. 290-311]. The standard analysis of Rayleigh flow concludes that increasing heat addition in such a flow inevitably leads to a sonic condition in the duct, referred to as *thermal choking*, distinct from the more common geometric or aerodynamic choking at the throat of a converging-diverging nozzle. Effectively, a supersonic (M > 1) flow in such a duct with increasing heat addition will tend to decelerate to sonic (M=1) and, likewise, a subsonic (M<1) flow will accelerate to sonic (M=1) condition. This is depicted by the traditional Rayleigh line on an *h-s* diagram as sketched by the full line in Figure 1. On the horizontal axis is the heat addition or increase in entropy or rise in total temperature – they all correspond to one another. The maximum entropy corresponds to the sonic condition (that is, the thermal throat). The vertical axis shows the enthalpy or static temperature. For the same total temperature (horizontal axis), higher static temperatures correspond to lower Mach numbers and vice versa. Thus, the upper segment of the Rayleigh line represents subsonic flow and the lower one supersonic flow (as marked in Figure 1). From a heat transfer point of view, the point $s_{max}$ represents the theoretical maximum heat that can be added to a Rayleigh flow until the thermal choking limit is attained.

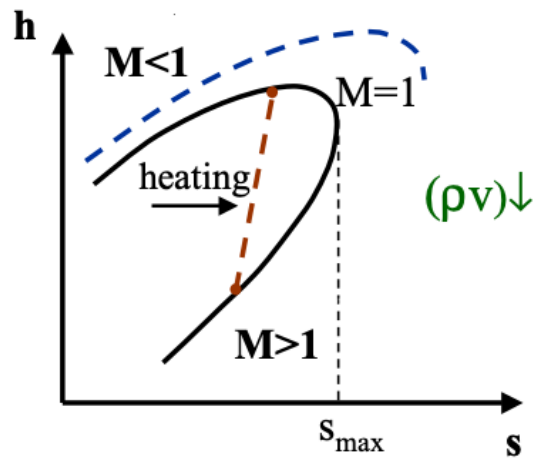

**Fig. 1 Traditional h-s diagram of Rayleigh line in constant-area, frictionless duct; dashed curve is Rayleigh line at lower mass flow rate. (Adapted from http://seitzman.gatech.edu/classes/ae3450/rayleighflow_two.pdf)**

In practice, flow in ducts with heat addition, such as combustors, will feature area change, friction and mass addition due to fuel injection. While these effects may be relatively negligible in subsonic combustors such as those in gas turbine engines, for a supersonic combustor such as in a scramjet engine they may be quite significant. Since the acceleration of the flow in an area-varying duct with friction will also depend on these factors, the Rayleigh line in Figure 1 will be suitably altered. Unfortunately, analytical solutions for heat addition in an arbitrary duct with mass injection, area change and friction have not been so forthcoming; the work in a recent paper [2] is a welcome development. Alternatively, one must numerically solve the differential equations for a quasi-one-dimensional flow [3] in a duct which allows all the effects mentioned above to be included. However, for conceptual analysis and preliminary design, an analytical solution or a zero-dimensional model amenable to rapid analysis is very much desirable. In particular, the conditions under which a thermally choked flow may occur in such an arbitrary combustor duct is of great interest and may be used to draw useful conclusions about the operability of the combustor. In heat-transfer terms, the problem in case of a variable-area duct needs to be slightly revised — here, instead of a maximum possible value of heat transfer limited by the thermal choking state, for different values of heat addition, thermal choking may occur at different locations in the duct (corresponding to different area ratios) or not at all. Hence, we have something like an inverse problem — given the heat transfer state (heat addition due to combustion and heat loss at the walls), identify the location of the thermal choking condition.

The flow and thermodynamic state of a thermally choked combustor when heat is further added is also a critical issue. In this case, according to standard Rayleigh theory [1], the mass flow rate in the duct will be reduced and a thermally choked state will persist. On the *h-s* diagram of Figure 1, the Rayleigh line with reduced mass flow rate is shown as the dashed curve which is shifted to the right vis-à-vis the original Rayleigh line. The sonic point on the dashed line now occurs at a larger value of s (corresponding to a higher heat addition value). However, in practice, the mismatch between the choked mass flow rate in the combustor and the mass inflow rate can lead to a build-up of pressure in the combustion chamber which can lead to the inlet shock system being ejected out of the inlet – a phenomenon known as *unstart* [4, pp. 350-373]. Due to flow spillage, the mass inflow rate of an "unstarted" inlet is reduced and this lower mass inflow rate can then match the reduced choked mass flow rate in the combustor. Thus, the combustor may remain operational but the efficiency of the system (intake plus combustor) is in question. For a large part, the analysis of thermal choking in duct flows and in combustion chambers has been in the steady state; investigations of transient phenomena have primarily been experimental [5]. The transient processes leading to unstart (or its unsteady version called *buzz*) arising out of a choked state in a ram combustor have been investigated numerically [6]. A similar numerical approach has been followed for studying transients arising in case of a pure ramjet [7] and a Dual Combustor Ramjet (DCR) combustor [8].

In general, thermal choking is not desirable in either a pure ramjet or a pure scramjet combustor where, as discussed earlier, it is likely to result in unstart and drastically affect the performance of the engine. In a dual-mode combustor, on the other hand, thermal choking is essential for the engine to operate in the ram mode [9]. A dual-mode combustor (DMC), as sketched in Figure 2, operates as a scramjet at higher Mach numbers. Fuel is injected at the head of a nearly constant-area combustor duct which is followed by a diverging nozzle downstream. Upstream of the combustor is the inlet and a short isolator segment. However, at lower Mach numbers, fuel is injected at the head of the divergent nozzle which now functions as a subsonic combustor and the DMC works like a ramjet. The nearly constant-area segment of the scramjet combustor now acts like an isolator shielding the inlet from the effects of the high ram combustion chamber pressure. The subsonic flow in the ram combustor must transition to a supersonic flow by the nozzle exit in the absence of a geometric throat in the nozzle. This is achieved by inducing a thermal throat in the nozzle where the flow is thermally choked. Downstream of the thermal throat, supersonic flow can expand to higher Mach numbers in the diverging duct. Thus, analysis of thermal choking is vital for both DMC and scramjet combustor, but for contrary reasons – in the DMC it is essential in order to ensure ram-mode operation, whereas in the scramjet it is necessary to prevent performance degradation. The issues of thermal choking, dual mode operation and mode transition in scramjet engines continue to be of current interest [10-12].

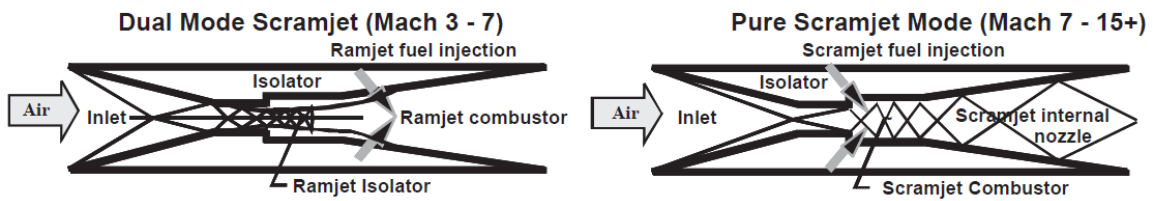

**Fig. 2  Schematic of dual-mode combustor in ram-mode (left panel) and scram-mode (right panel) operation [13, p. 183].**

In the rest of this paper, the zero-dimensional analysis of flow in an arbitrary duct with area variation, friction and mass addition is formulated and solved. After due validation, the Rayleigh line for varying area ratio ducts is generated. Likewise, the Rayleigh flow with friction is studied and the combined effect of friction and diverging duct equivalent to frictionless flow in a constant-area duct is evaluated. Attention is then focused on a model scramjet combustor that has been studied in the open literature featuring area change, wall friction, heat addition due to combustion and wall heat loss. After tuning the parameters in the zero-dimensional analysis to match the scramjet results in the literature, the analysis is used to investigate the occurrence of thermal choking in the scramjet combustor duct. Finally, the zero-dimensional analysis is used to analyze the conditions for occurrence of thermal choking in a dual-mode combustor operating in the ram mode. The combinations of fuel-air equivalence ratio and combustor area divergence for which a sonic state can be supported are calculated and presented.

## II. Heat Addition in Arbitrary Duct

Consider an arbitrary duct with inlet station labeled '1' and outlet station labeled '2'. For a zero-dimensional analysis, the precise area variation of the duct is of no importance; only the inlet and outlet areas matter. The equations for a zero-dimensional analysis of compressible flow in a duct are those for conservation of mass, momentum and energy along with the equation of state. Along the lines of Ref. [2], these are written as:

$$\rho_2 U_2 A_2 = (1+f)\rho_1 U_1 A_1 \tag{1}$$

$$p_2 A_2 + \rho_2 U_2^2 A_2 = p_1 A_1 + \rho_1 U_1^2 A_1 + p_1(A_2 - A_1) - \left(\frac{1}{2}\right)\rho_1 U_1^2 C_f A_w \tag{2}$$

$$\frac{\gamma_2 R_2}{\gamma_2 - 1}(T_2 - T_1) + \left(\frac{1}{2}\right)(U_2^2 - U_1^2) = \frac{\dot{Q}_c + \dot{Q}_w}{(1+f)\rho_1 U_1 A_1} \tag{3}$$

$$p_2 = \rho_2 R_2 T_2; p_1 = \rho_1 R_1 T_1 \tag{4}$$

Defining the following non-dimensional variables:

$$r = \frac{\rho_2}{\rho_1}; u = \frac{U_2}{U_1}; AR = \frac{A_2}{A_1}; \bar{A}_w = \frac{A_w}{A_1}; \bar{R} = \frac{R_2}{R_1}; \bar{\dot{Q}}_c = \frac{\dot{Q}_c}{\rho_1 U_1^3 A_1}; \bar{\dot{Q}}_w = \frac{\dot{Q}_w}{\rho_1 U_1^3 A_1} \tag{5}$$

Equations (1) – (3) can be represented as follows, where Eq. (4) has been incorporated:

$$r.u.AR = 1+f \tag{6}$$

$$r.u^2.AR.\frac{R_2 T_2}{U_2^2} + r.u^2.AR = AR.\frac{R_1 T_1}{U_1^2} + 1 - \left(\frac{1}{2}\right)C_f \bar{A}_w \tag{7}$$

$$\frac{\gamma_2}{\gamma_2 - 1}\left(u^2.\frac{R_2 T_2}{U_2^2} - \bar{R}.\frac{R_1 T_1}{U_1^2}\right) + \left(\frac{1}{2}\right)(u^2 - 1) = \frac{\bar{\dot{Q}}_c + \bar{\dot{Q}}_w}{1+f} \tag{8}$$

Clearly the density ratio $r$ and the velocity ratio $u$ are two of the variables in Eqs. (6) – (8). The choice of the third variable is crucial – this is selected as the outlet Mach number $M_2$ since it is a key parameter for thermal choking and the Rayleigh flow is also best parameterized in terms of $M_2$. With the following additional definitions:

$$M_1^2 = \frac{U_1^2}{\gamma_1 R_1 T_1}; M_2^2 = \frac{U_2^2}{\gamma_2 R_2 T_2}; \bar{\gamma} = \frac{\gamma_2}{\gamma_1} \tag{9}$$

Equations (6) – (8) can be written in the form below:

$$r.u.AR = 1+f \tag{10}$$

$$r.u^2.AR.\left(\frac{1}{\gamma_2 M_2^2} + 1\right) = \left(AR.\frac{1}{\gamma_1 M_1^2} + 1\right) - \left(\frac{1}{2}\right)C_f \bar{A}_w \tag{11}$$

$$\frac{1}{\gamma_2 - 1}\left(u^2.\frac{1}{M_2^2} - \bar{R}.\bar{\gamma}.\frac{1}{M_1^2}\right) + \left(\frac{1}{2}\right)(u^2 - 1) = \frac{\bar{\dot{Q}}_c + \bar{\dot{Q}}_w}{1+f} \tag{12}$$

Equations (10) – (12) are a set of three nonlinear algebraic equations in the three variables – $r$, $u$, $M_2$, of the form:

$$F_i(r, u, M_2) = 0; i = 1,\ldots,3 \tag{13}$$

Equation (13) can be solved in the most general case, for example, by using the MATLAB® solver *fsolve*.

### A. Validation

First of all, the solution of Equations (10) – (12) is validated against a result presented in Ref. [2] for the inlet conditions and parameters presented in Table 1. Results are generated for the case of a supersonic flow in a duct with inlet Mach number $M_1 = 2.5$ and other parameters as given in Table 1. Five different cases of area ratio ($AR$) are considered and the non-dimensional heat addition at the wall, $\bar{Q}_w$, is varied. The equations for generalized Rayleigh flow in the form of Eq. (13) are solved for the three variables. A check is placed to catch the iteration at which the outlet Mach number attempts to transition past the sonic state, $M_2 = 1$. In that case alone, the outlet Mach number of $M_2 = 1$ is given as one of the input states and the non-dimensional heat addition at the wall is calculated instead as the third unknown variable.

**Table 1 Inlet conditions and parameters for validation case**

| Inlet condition/Parameter | Value | Inlet condition/Parameter | Value |
|---|---|---|---|
| $\bar{\gamma}$ | 0.91 | $f$ | 0 |
| $\gamma_1, \gamma_2$ | 1.32, 1.20 | $C_f$ | 0 |
| $\bar{R}$ | 1.0 | $\bar{Q}_c$ | 0 |
| $\bar{A}_w$ | 10.0 | $M_1$ | 2.5 |

For the validation, the outlet Mach number $M_2$ is the parameter of interest, especially the condition $M_2 = 1$ which corresponds to a thermally choked flow at the outlet. These results are shown in Figure 3 with the plot from Ref. [2] shown alongside for comparison. The only difference between these plots is the variable of choice on the horizontal axis – the two are entirely equivalent except for a difference in scale. They can be completely matched by scaling down the horizontal axis in our plot by a factor of 0.7. As can be seen from Figure 3, the results in the two plots are totally identical. Further the calculation in Ref. [2] was checked against the result computed by the MATLAB® solver *fsolve* in one instance and each element of the 3-vector in Eq. (13) was found to be accurate to *1e-16*.

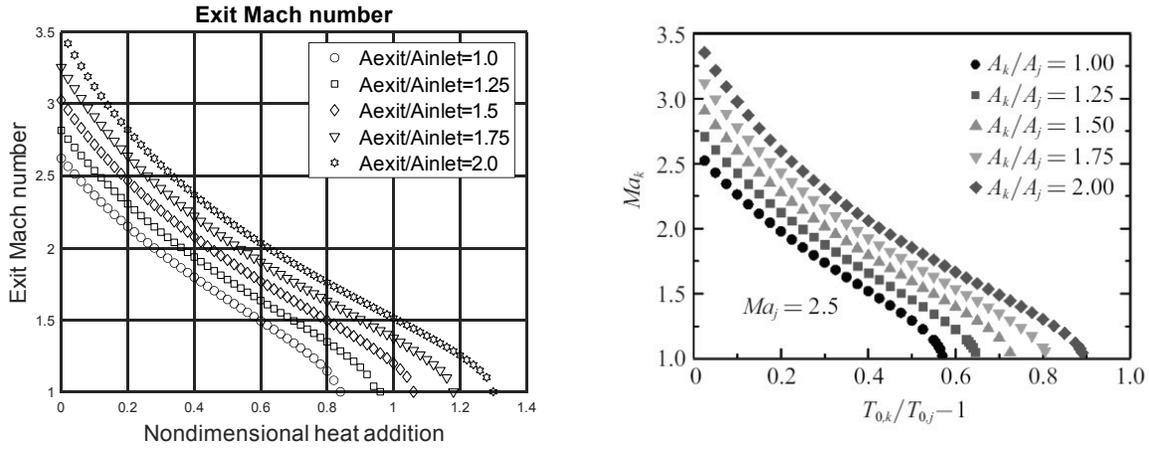

**Fig. 3 Validation case for Rayleigh flow in an area-varying duct; plot of outlet Mach number $M_2$ against the non-dimensional heat addition (left plot – this paper, right plot – Ref. [2]).**

The values of non-dimensional heat addition for which thermal choking occurs for each area ratio case are listed in Table 2. These values are the intercepts on the horizontal axis in the left plot of Figure 3 corresponding to each area ratio case. As discussed earlier, this can be considered as the "forward" heat transfer solution — the non-dimensional heat addition that leads to thermal choking is a measure of the limit of entropy increase to the Rayleigh flow beyond which this level of choking mass flow can no longer be maintained.

**Table 2 Maximum value of heat addition leading to thermal choking for each area ratio case**

| Area Ratio | Non-dimensional heat addition at thermal choking state |
| --- | --- |
| 1.0 | 0.8298 |
| 1.25 | 0.9425 |
| 1.5 | 1.0582 |
| 1.75 | 1.1769 |
| 2.0 | 1.2986 |

One minor anomaly that may be commonly observed in both plots of Figure 3 is that the limiting value on the vertical axis, corresponding to no heat addition, does not match the isentropic flow solution. This happens because the specific heat ratio at the outlet $\gamma_2$ has been taken in Ref. [2] to be the lower "hot flow" value of 1.2 as given in Table 1 irrespective of the amount of heat addition. For purposes of validation, the same selection was retained in producing the left plot in Figure 3. This discrepancy is now resolved by scheduling $\gamma_2$ between the "cold flow" and "hot flow" values depending on the value of the heat addition. The "corrected" plots of the outlet Mach number $M_2$, velocity ratio $u$, and density ratio $r$, are shown in Figure 4. Now, for instance, in the constant-area duct case, the outlet Mach number can be verified to be the same as the inlet Mach number when there is no heat addition (friction has been neglected in the solutions so far). There is no change in the intercepts on the horizontal axis – the thermal choking states are not altered between Figure 3 and Figure 4.

**B. Rayleigh lines with varying area ratio**

It is immediately obvious from Figure 4 that a larger amount of heat can be added before the thermal choking state is reached as the area ratio increases. In supersonic flow, increasing area tends to increase the Mach number while heat addition tends to push the Mach number towards sonic – the two effects counteract each other. For this reason, scramjet combustors, at least partly, are given a diverging section.

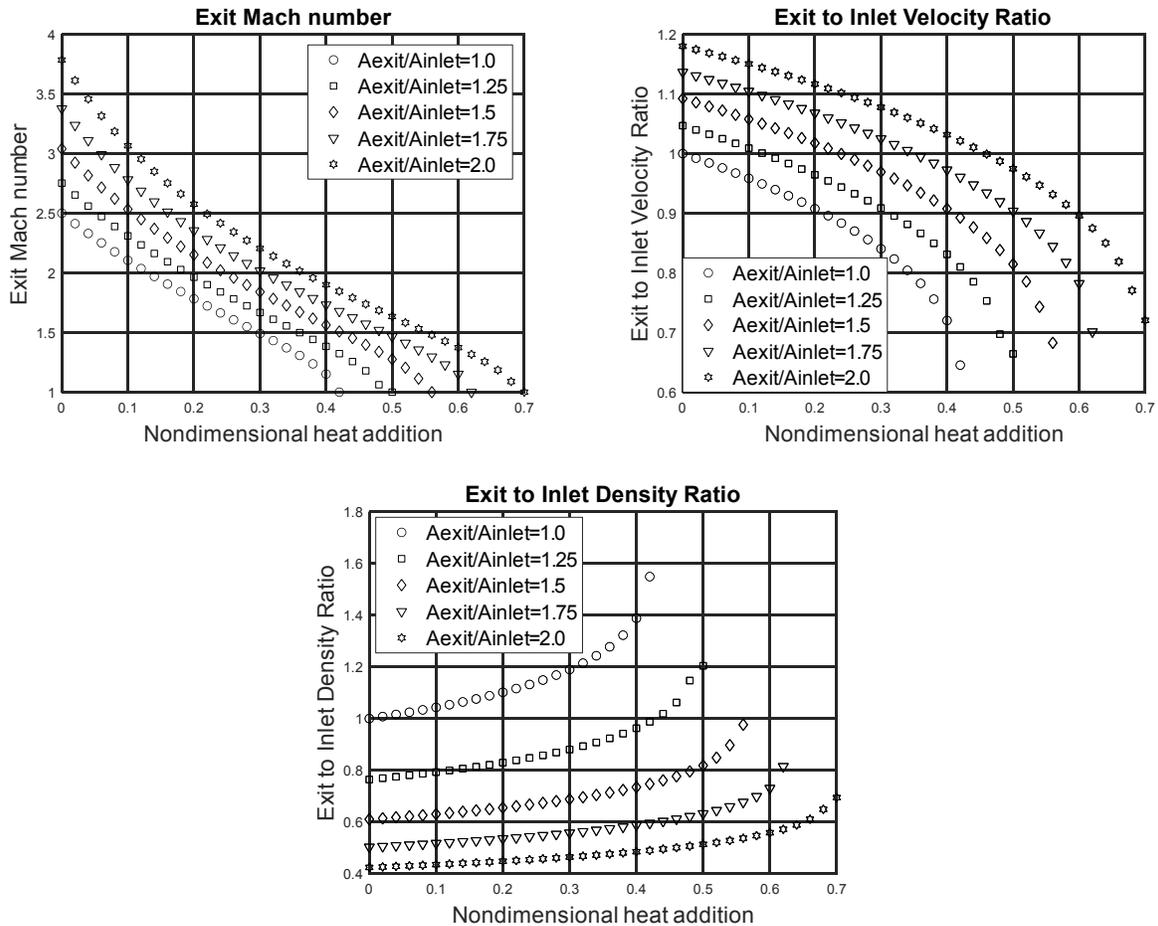

**Fig. 4 Variables in Rayleigh flow in area-varying duct with the "hot gas" specific heat ratio corrected.**

Carrying the calculations in Figure 4 further, Rayleigh lines for different area ratio cases are shown in Figure 5. The standard Rayleigh line for a constant-area duct is indicated by the full black line; the others are lines for diverging ducts with different area ratios. It is seen that for a divergent duct the Rayleigh line shifts to the right. Compared to the case in Figure 1 where heat addition in a constant-area duct at the choking state caused a shift to the dashed line at lower mass flow rate, in Figure 5 the same mass flow rate can be maintained while shifting the Rayleigh line to the right by increasing the area at the thermal throat. The result in Figure 5 therefore correlates perfectly with the traditional understanding in Figure 1 — additionally, variable-area ducts have the freedom to adjust the location of the thermal choking condition to match the given heat transfer state for the same mass flow rate. Note that, flow being supersonic, only the lower segment of the Rayleigh line (refer Figure 1) is seen in Figure 5. To summarize, in constant-area ducts at a thermally choked state, either heat transfer is limited or mass transfer through the duct is reduced; on the other hand, variable-area ducts can accommodate various heat transfer levels for a given mass transfer state as long as a location for thermal choking with the appropriate area ratio is available.

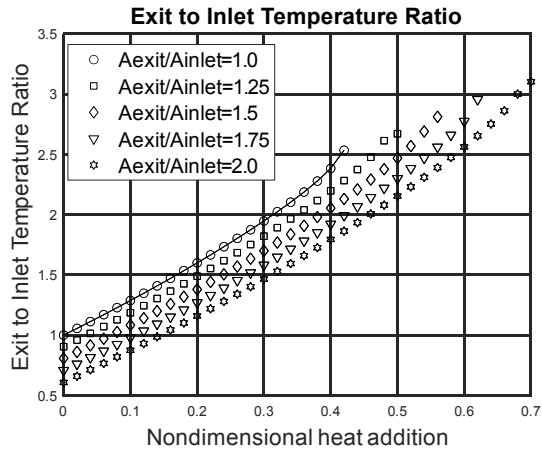

**Fig. 5  Rayleigh lines for different duct area ratio cases.**

## C.  Rayleigh lines with friction

Practically, duct flows must account for frictional effects. In the zero-dimensional formulation here, these effects are lumped into a single friction factor $C_f$. It is of interest to observe the effect of increasing frictional losses in the duct on the Rayleigh lines. The results for frictionless flow and flow with four different levels of $C_f$ for a constant-area duct are shown in Figure 6. As expected, the effect of increased friction with no heat addition is to reduce the outlet Mach number as seen from the points on the vertical axis in the left panel of Figure 6. The intercepts on the horizontal axis of the same panel indicate that in the presence of friction lesser heat may be added to the duct before a thermally choked state is obtained. This is reasonable as both friction and heat addition here have a similar effect – that of slowing down the flow. The change in the Rayleigh lines with increasing friction is depicted in the right panel of Figure 6 – the lines move to the left relative to the traditional one without friction (full black line). In fact for the largest value of $C_f$ in Figure 6, friction alone reduces the inlet Mach number of 2.5 to sonic at the outlet, with no scope for any heat addition at all.

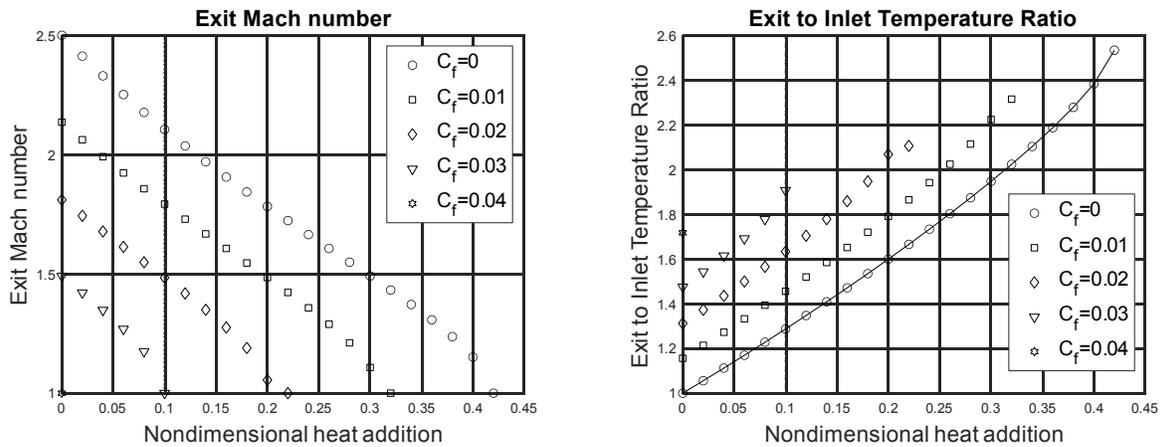

**Fig. 6  Effect of varying frictional loss on the maximum heat addition in a constant-area duct and the Rayleigh line.**

Clearly, from Figure 5 and Figure 6, the effects of area divergence and friction on the Rayleigh line are in opposing senses. The gains due to duct divergence may be lost when frictional effects are considered. Figure 7 reproduces the plots of Figure 6 with an additional set of results for the case of $C_f = 0.01$ and area ratio of 1.4 superimposed. The superimposed data may be seen to fall exactly on the original Rayleigh line. In other words, the advantage in terms of higher heat addition gained by providing an area ratio of 1.4 is precisely negated by the frictional effect of $C_f = 0.01$. This suggests that in scramjet combustors the divergence of the duct in reality may only, more or less, offset the losses due to frictional effects and may not allow for any significant additional heat input.

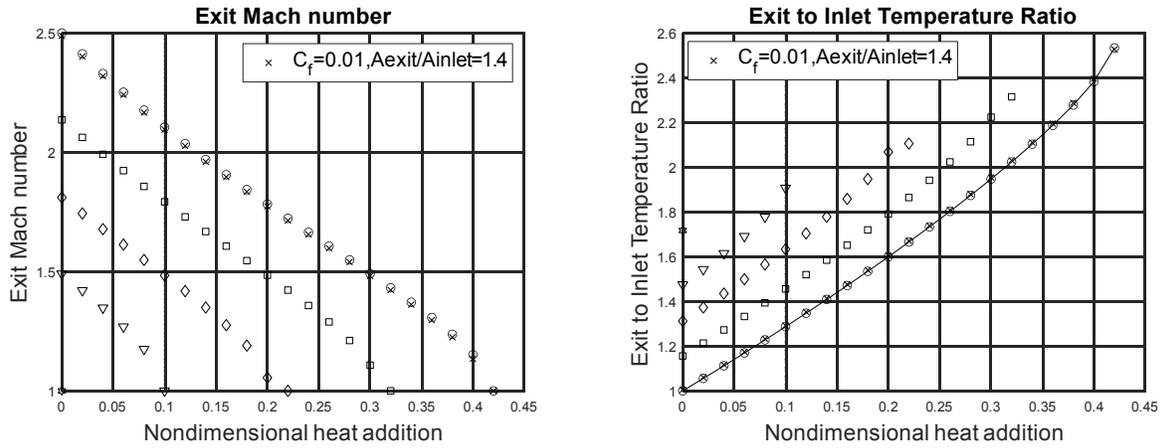

**Fig. 7 Figure 6 with a combined case of friction and diverging duct area superimposed (crosses).**

## III.                               Application to a Scramjet Combustor

We now apply the zero-dimensional analysis developed so far to investigate the possibility of thermal choking in a scramjet combustor. The combustor geometry taken from Ref. [3] is as shown in Figure 8. It is a rectangular duct of fixed width; there is an initial segment of constant area, followed by a divergent segment where the upper wall diverges at an angle of 8.39 deg. Details of the combustor geometry and the inlet flow conditions are listed in Table 3. Results from a quasi-one-dimensional analysis of the geometry in Figure 8 have been presented in Ref. [3]. Heat transfer at the wall and frictional loss have been considered in the quasi-one-dimensional analysis along with combustion of Jet-A fuel in the duct; hence all three effects – heat addition/loss, viscous/ friction loss and area change are present in this problem of a model scramjet combustor.

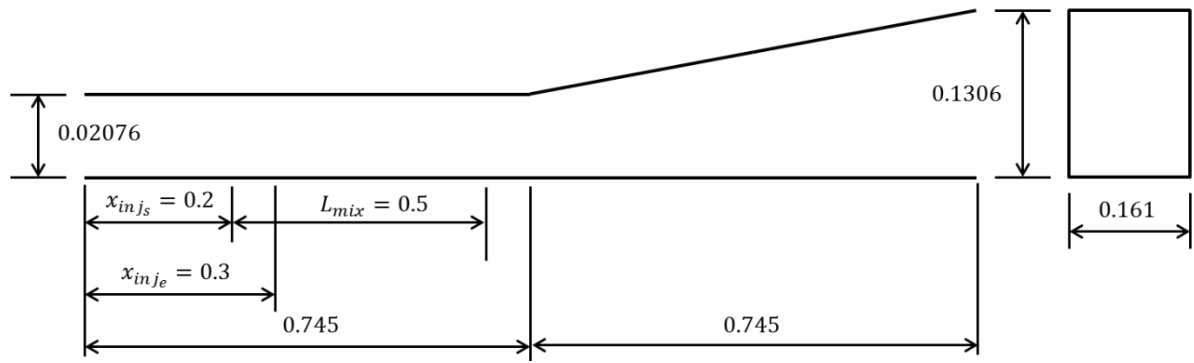

**Fig. 8 Geometry of scramjet combustor used for thermal choking analysis – not to scale (from Ref. [3]).**

Table 3 Flow and geometry specifications for the scramjet combustor model in Figure 8

| Quantity | Value |
|---|---|
| Inlet static pressure | 1.519 atm |
| Inlet static temperature | 1335 K |
| Inlet Mach number | 2.563 |
| Inlet air specific heat ratio | 1.318 |
| Inlet air mass flow rate | 2.435 kg/s |
| Injected fuel mass flow rate | 0.152 kg/s |
| Fuel-air equivalence ratio | 0.929 |
| Duct wall temperature | 1200 K |
| Fuel mixing efficiency | 90% |
| Fuel injection start station | 0.2 m |
| Fuel injection end station | 0.3 m |
| Fuel mixing length | 0.5 m |
| Length of constant-area combustor segment | 0.745 m |
| Length of diverging-area combustor segment | 0.745 m |
| Diverging-area segment expansion angle | 8.39 deg |
| Combustor width | 0.161 m |
| Fuel | Jet-A (C12H23) |
| Fuel density | 855 kg/m$^3$ |
| Calorific value of Jet-A fuel | 42.8 MJ/kg |

Key results from Ref. [3] for this model scramjet combustor have been reproduced in Figure 9. The Jet-A fuel is injected between axial stations at 0.2 m and 0.3 m. Mixing efficiency of 90% is set; hence 10% of the fuel is never mixed and remains unreacted until the exit station at 1.49 m. A mixing length is prescribed leading to mixing for 90% of the fuel being completed by axial station at 0.7 m. Reaction is initiated thereafter and all the mixed Jet-A fuel is combusted by axial station 0.84 m.

In Figure 9, there is an initial drop in Mach number at locations prior to the start of fuel injection; this is attributed to viscous effects in the duct. There is a steeper drop in Mach number at the fuel injection stations due to liquid fuel being introduced transversally. Subsequently, Mach number continues to fall through the fuel mixing length due to frictional effects. Note the fall in Mach number from 2.563 at inlet to around 1.6 at the point where reaction is initiated. Mach number then falls more steeply in the constant-area duct due to heat addition from combustion. After a brief discontinuity at the point where the duct begins to diverge, Mach number falls over the rest of the reaction length falling to a low of 1.08 at axial station 0.84 m – the point where all the mixed fuel has effectively been combusted. Thereafter, Mach number rises in the diverging duct recovering to a value very nearly the same as the inlet value.

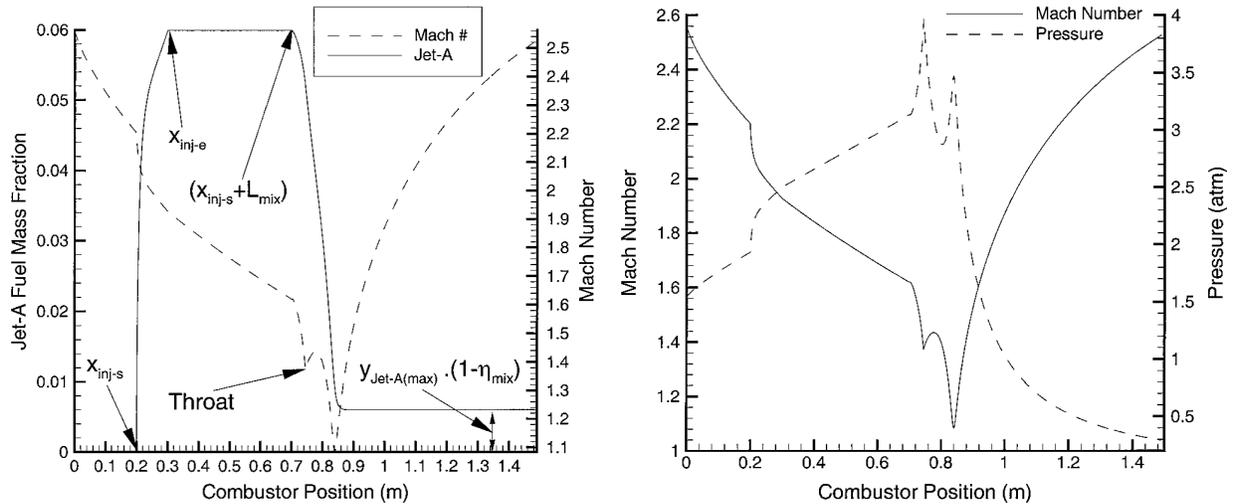

**Fig. 9  Quasi-1D analysis results for the scramjet combustor geometry in Figure 8 (taken from Ref. [3]).**

### A. Zero-dimensional analysis – Estimation of unknown parameters

The analysis presented here may be used in one of two complementary ways: Firstly, we can select a particular station '2' as the outlet for the analysis. Then, knowing $A_2$, one may determine the possibility of thermal choking at station '2', or find values of certain parameters for which thermal choking occurs at that location. Alternatively, one may let the location of station '2' be initially undetermined but select $M_2 = 1$ and then solve for $A_2$, and hence find the location in the duct where thermal choking takes place. Both approaches are useful in practice, as demonstrated below.

For the purposes of the zero-dimensional analysis of the scramjet model in Figure 8, the station '1' is taken as the inlet at x=0 and station '2' at x=0.84 m. The area ratio between these two stations is *AR=1.675*. The non-dimensional wetted area over this length is $\bar{A}_w = 91.767$. In case of cold flow, the Mach number at station '2' is estimated by extrapolation of the data in Figure 9 to be 1.525. Then assuming both heat added due to combustion and heat loss at the wall to be zero in case of cold flow, the zero-dimensional analysis is carried out with $M_2 = 1.525$ and the average friction coefficient for the duct is back-calculated. In this manner, the friction coefficient is estimated as $C_f = 0.0049$. The heat added due to combustion may be calculated since the calorific value of fuel, fuel mass flow rate and mixing efficiency are given in Table 3. Thereby, the non-dimensional heat added due to combustion of Jet-A fuel in the scramjet is found to be $\bar{Q}_c = 0.7219$. Finally, the non-dimensional heat addition at the wall is to be estimated. From Figure 9, for the hot flow case, the exit Mach number is known to be 1.082. With the above values of $C_f = 0.0049$ and $\bar{Q}_c = 0.7219$, the zero-dimensional analysis is once again carried out, this time with $M_2 = 1.082$ and the non-dimensional heat addition at the wall is back-calculated. This yields the result $\bar{Q}_w = -0.552$ suggesting that heat is being lost to the surroundings at the wall. Thus, the parameters of the zero-dimensional analysis have been 'tuned' to match the results of the model scramjet in Ref. [3]. The zero-dimensional analysis code can now be used to investigate the phenomenon of thermal choking in the scramjet model of Figure 9.

**B. Zero-dimensional analysis – Thermal choking**

Two parameters are used to study the possibility of thermal choking in the scramjet model under investigation. First, keeping the fuel mass flow rate unchanged, the wall heat transfer is altered so as to create a state of thermal choking at the station '2' at x=0.84 m. This is easily done since the zero-dimensional model already has an option to specify $M_2 = 1$ and compute the non-dimensional wall heat transfer as the third variable (in addition to the density and velocity ratios). In this manner, a thermally choked state at the outlet station '2' is found for $\bar{Q}_w = -0.5211$. This is a mere 5.6% change in $\bar{Q}_w$ from its baseline value of $\bar{Q}_w = -0.552$. One way by which this decrease can be engineered is to set a lowered value of the adiabatic wall temperature.

Alternatively, the fuel mass flow rate injected into the combustor may be changed, which changes the non-dimensional heat added due to combustion $\bar{Q}_c$. Then the curves of outlet Mach number with varying non-dimensional heat addition can be calculated for each setting of the fuel mass flow rate as was done for different values of $C_f$ in Figure 6. This is displayed in Figure 10 which also shows the Rayleigh lines traced for the different fuel injection rate cases. Expectedly, at higher fuel mass flow rates (below equivalence ratio of 1), the increased $\bar{Q}_c$ requires a larger negative value (greater wall heat loss) to maintain a thermally choked state at the exit station '2' for the same value of air mass flow in the combustor duct. (The change in gas flow rate in the duct due to the change in fuel injection rate is negligible.)

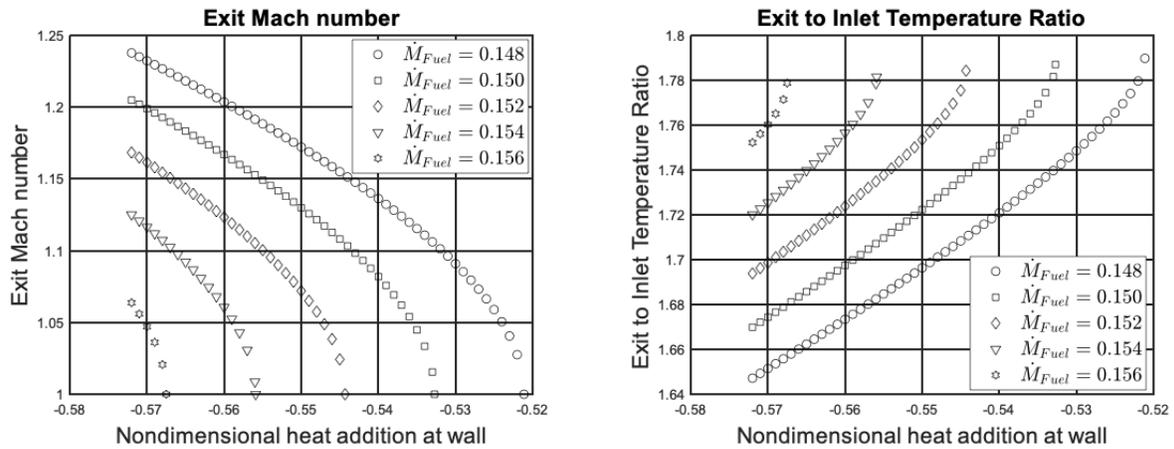

Fig. 10    Non-dimensional wall heat transfer condition for thermal choking at outlet station '2' for the scramjet combustor model for different fuel injection rates.

One way of interpreting the results in Figure 10 is to deduce the change in the non-dimensional wall heat addition per unit change in the fuel mass injection rate such that a thermally choked state is maintained at the exit station '2.' These data are extracted from Figure 10 and cross-plotted in Figure 11. Interestingly, the variation is linear with a slope of -0.1725 per kg/s of fuel mass injection rate. By syncing the changes in the fuel mass injection rate to the wall heat transfer in this ratio, a constant state of thermal choking can be maintained in case of this scramjet combustor model.

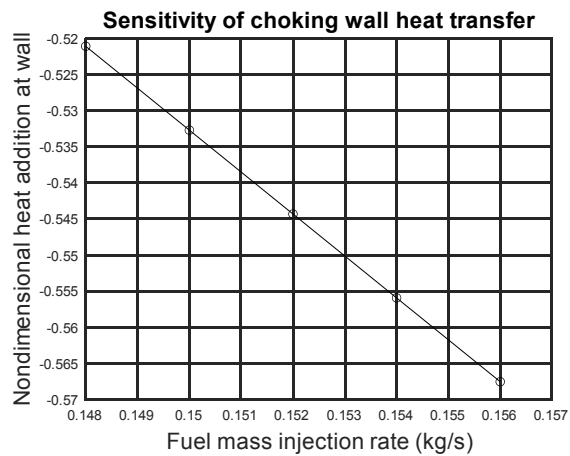

Fig. 11    Sensitivity of non-dimensional wall heat addition/loss to maintain a thermally choked state at the outlet station to changes in fuel mass injection rate.

While the occurrence of thermal choking may not be a welcome phenomenon in case of a scramjet engine, for the dual mode combustor DMC engine in the ram mode it is a necessity, as we investigate next.

IV.                          Application to Dual-mode Ramjet Combustor

Figure 12 is a schematic of an axisymmetric dual-mode scramjet combustor geometry with the key dimensions marked. The length has been marked into four main segments: the first is a constant-area segment (inner diameter: φ215, outer diameter: φ237) for the fuel injection station for scram-mode operation. The second segment is a diverging one (inner diameter: φ215, outer diameter: φ237 – φ245) with a cavity for flame holding at the beginning and a step at the end. This acts as a supersonic combustor in the scram mode and as an isolator in the ram mode. The following segment is a constant-area section (inner diameter: φ215, outer diameter: φ263), followed by a divergent one (inner diameter: φ215, outer diameter: φ263 – φ283). The constant-area section is where fuel is injected for ram mode operation followed by a cavity for flame holding. The divergent section acts as a subsonic combustor in the ram mode and as an extension of the supersonic combustor in the scram mode. This is followed by a nozzle segment.

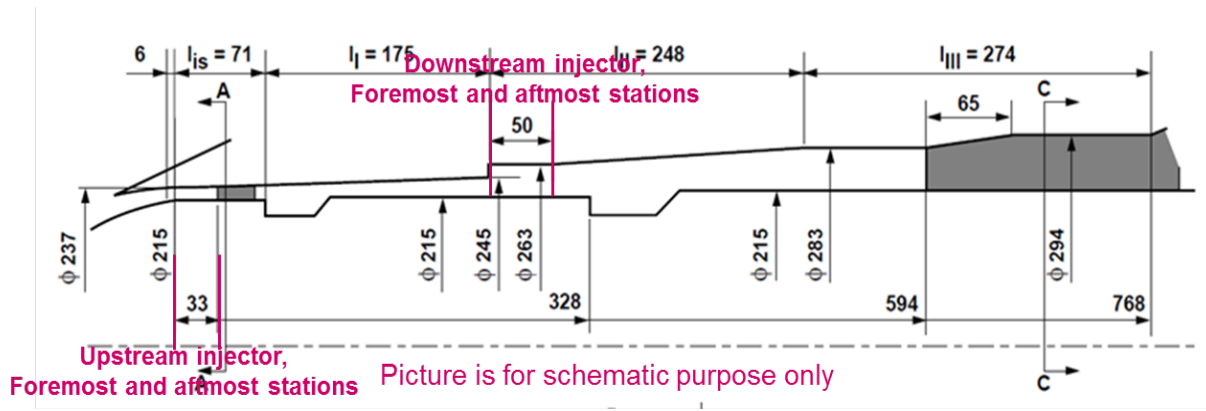

**Fig. 12**     Schematic of a dual-mode combustor showing key dimensions, all mm. (source: unknown)

In the ram mode, it is the third segment that acts as the ram combustor that is of interest. The flow and geometric features of this ram combustor are presented in Table 4. This corresponds to a Mach 6 flight of the vehicle at an altitude of 27 km. The most conservative case is considered with non-dimensional wall heat addition $\bar{Q}_w$ set to zero, as well as the wall friction coefficient $C_f = 0$. Both heat loss at the wall and friction serve to decelerate the subsonic flow making it difficult to accelerate to the thermal choking state of Mach 1. Thus, in the present calculations, the heat released due to combustion pushes the flow towards the sonic state and only the area divergence holds it back from thermal choking. In the presence of wall heat loss and friction, a lesser degree of area divergence would yield the same result. That is, the present result corresponds to the maximum area ratio required for a thermally choked state to occur at the given conditions.

**Table 4 Flow and geometry specifications for the dual-mode combustor model in Figure 12**

| Quantity | Value |
| --- | --- |
| Combustor inlet area, $A_1$ | 0.0180 m² |
| Combustor outlet area | 0.0266 m² |
| Non-dimensional wetted area (combustor), $\bar{A}_w$ | 22.6556 m² |
| Combustor inlet Mach number, $M_1$ | 0.4 |
| Combustor inlet static pressure, $p_1$ | 681.3088 Pa |
| Combustor inlet static temperature, $T_1$ | 1468.8 K |
| Combustor inlet velocity, $U_1$ | 298.1 m/s |
| Fuel mixing efficiency | 100% |
| Combustor inlet air mass flow rate | 0.0087 kg/s |
| Fuel | Jet-A (C12H23) |
| Fuel density | 855 kg/m³ |
| Calorific value of Jet-A fuel | 42.8 MJ/kg |
| Inlet air specific heat ratio, $\gamma_1$ | 1.318 |
| Outlet gas specific heat ratio, $\gamma_2$ | 1.2 |
| Ratio of gas constant at outlet to inlet, $\bar{R}$ | 1 |
| Non-dimensional wall heat addition, $\bar{\dot{Q}}_w$ | 0 |
| Wall friction coefficient, $C_f$ | 0 |

Calculations are carried out for five different values of fuel-air equivalence ratio (ER) between 0.4 and 1.0 and reported in Figure 13. The curve for each value of ER intersects the $M_2 = 1$ axis for a certain value of the Area Ratio thereby giving the location of the thermally choked condition in the ram combustor segment. Beyond the thermal choke state, the ram flow transitions into a supersonic flow that expands in the nozzle to produce thrust. As expected, for the lower values of ER the heat addition is just insufficient to push the flow in the ram combustor to a sonic state, even assuming a constant-area duct. For ER in the range 0.7 to 1.0, thermal choking is seen at increasing area ratio with increasing ER — for different values of heat addition, the location of the thermal choke point is different as given by the area ratio corresponding to exit Mach number of 1 (that is, the thermal choking location moves downstream in the duct from area ratio of 1.02 to 1.15, approximately, as the ER, and hence the heat addition, is increased from 0.7 to 1.0). This is a solution of the "inverse" heat transfer problem for Rayleigh flow mentioned earlier. Note that, when solving the "inverse" problem, the outlet station '2' is not a given, fixed location, but is variable and is obtained as part of the solution.

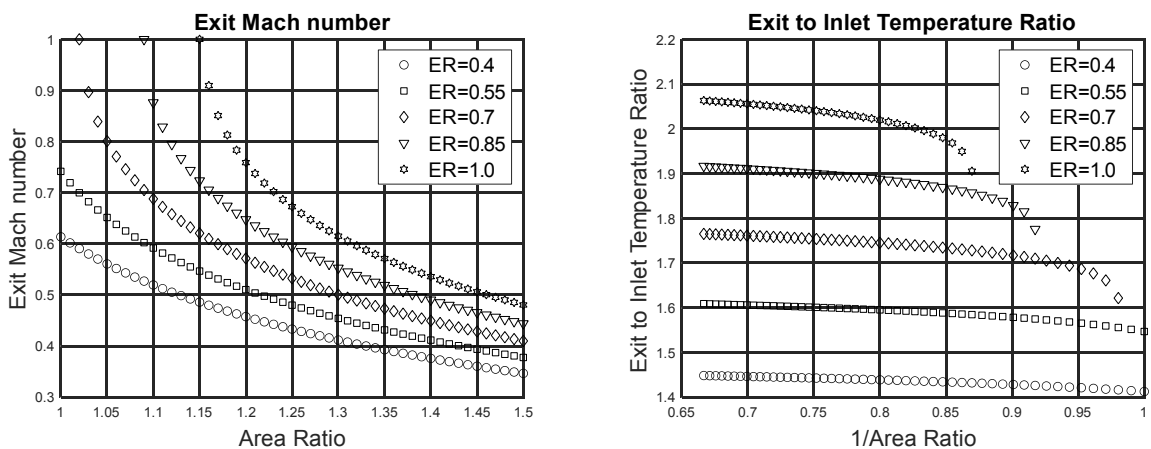

**Fig. 13    Thermal choking states in the dual-mode combustor geometry of Figure 12 for various equivalence ratio cases.**

Figure 13 also shows the static temperature ratio (equivalently, the enthalpy ratio) for the different ER cases. This figure is plotted with the inverse of Area Ratio on the X-axis to give each curve an appearance similar to the subsonic Rayleigh line. It can be seen that in the three highest ER cases the curve tends to an infinite slope towards the Mach=1 condition as a typical Rayleigh line would. Results such as the one in Figure 13 may be used to design the dual-mode combustor or to analyze its performance or to decide its operating condition (e.g., choice of equivalence ratio). Wall heat loss and friction, once estimated, can be easily included in the calculations to get more realistic results. Finally, it must be noted that the high pressure prevailing in the ram combustor usually results in a region of separated flow; hence the effective flow area will usually differ from the geometric one. The analysis can be modified to account for this effect if empirical or approximate models for the area covered by the separated flow are provided.

# V. Conclusion

A zero-dimensional model has been used to analyze generalized Rayleigh flow in combustors with multiple parameters including heat addition due to combustion, frictional losses, varying duct area, wall heat loss, and varying fuel injection rates. The model has three variables – the duct density ratio, the velocity ratio, and the exit Mach number – which can be studied with the variation of any of the parameters listed above. The particular focus of this work has been the occurrence of a thermally choked state in the ramjet combustor. The model has been applied to a dual-mode combustor operating in the ram mode with Jet-A fuel where thermal choking is a necessary condition. Conditions for thermal choking at various duct locations and fuel-air equivalence ratio are evaluated. In particular, solutions to the "inverse" heat transfer problem have been demonstrated — given the duct mass flow and heat transfer state, locate the point of occurrence of thermal choking. The model is also used to study the possibility of thermal choking in a scramjet combustor where it is an undesirable phenomenon – in this case all the effects listed as parameters above are included as well as fuel mixing efficiency. The model developed here can be a useful tool for conceptual and preliminary design and analysis of combustor systems as demonstrated in this work. It may also be used in conjunction with a quasi-one-dimensional combustor code as a first estimate for the occurrence and possible location of the thermally choked state.